\documentstyle[preprint,prb,aps]{revtex}
\def\be{\begin{equation}}
\def\ee{\end{equation}}
\def\bc{\begin{center}}
\def\ec{\end{center}}
\begin{document}
\begin{titlepage}
\bc
{\Large \bf Step fluctuations and random walks}\\[5ex]
 
M. Bisani and W. Selke\\

Institut f\"ur Theoretische Physik, Technische Hochschule,
D--52056 Aachen, Germany \\
\ec

\vskip 1.5cm
\noindent{\Large \bf Abstract}\\[2.5ex]
The probability distribution $p(l)$ of an atom to
return to a step at distance $l$
from the detachment site, with a random walk in between, is exactly
enumerated. In particular, we study the dependence of $p(l)$ on step
roughness, presence of other reflecting or adsorbing steps,
interaction between steps and the diffusing atom, as well as
concentration of defects on the terrace neighbouring the step.
Applying Monte Carlo techniques, the time
evolution of equilibrium step
fluctuations is computed for specific forms of return probabilities.
Results are compared to previous theoretical and experimental
findings.\\[1cm]


\noindent
{\it Corresponding author:}\\
Prof. Walter Selke\\
Institut f\"ur Theoretische Physik, Technische Hochschule,
 D--52056 Aachen, Germany\\
Tel: 0049-241-80-7029; Fax: 0049-241-8888-188\\
e-mail: selke@physik.rwth-aachen.de\\
\end{titlepage}

{\Large \bf 1. Introduction}\\

In recent years, the dynamics of steps on crystal surfaces has
attracted much interest, both experimentally [1-6] and 
theoretically [7-11]. Experimentally, equilibrium step
fluctuations of isolated steps as well as steps on
vicinal surfaces have been studied extensively, following pioneering
STM measurements of Au(110) [1] and Cu(11n) [2] surfaces.
The step fluctuations are quantified by the correlation
function $G(t)$= $\langle (h(i,t)- h(i,0))^2 \rangle$, where $h(i,t)$
denotes the position (or displacement) $h$ of the
step at site $i$
and time $t$. Typically, the experimental data are
described by a power law, $G(t) \propto t^{\gamma}$, with
the dynamic exponent $\gamma$ being in between about 1/4 and 
about 1/2. Theoretically, distinct atomic mechanisms
driving the step dynamics have been identified, leading, indeed, at
long times, to power laws, with $\gamma$ being
1/4, 1/3, and 1/2 in the limiting cases of
step diffusion, terrace diffusion, and evaporation--condensation
kinetics, respectively. These
scenarios have been found in Langevin descriptions [7,9,10]
and confirmed, partly, in simulations on SOS models [7,11].\\ 

\noindent
In this article, step fluctuations will be analysed in
the framework of random walks of atoms detaching from
the (reference) step, diffusing on the neighbouring terrace, and
attaching again at the step at distance $l$ (in units of the
lattice spacing on the surface) from the detachment site. The
corresponding return probability distribution $p(l)$ is
determined for various situations, to analyse the influence
of the step roughness, of the presence of another reflecting or
adsorbing step separated by $d$ lattice spacings from
the reference step, of the interaction of the diffusing
atom with the steps due to, e.g., elastic forces, as well as
of the concentration of reflecting or adsorbing defects
on the terrace. The case of a straight step with
no obstacles on the terrace and without interactions
between the atom and the step has been considered
before, solving the related continuum diffusion equation, 
in the context of a Langevin theory, where $p(l)$
is called diffusion kernel [9]. Here, results on $p(l)$
will be then incorporated in Monte Carlo simulations of the step
fluctuations. We shall present simulational results for special
choices of return probabilities connecting
detachment and attachment step sites, including the three
limiting cases mentioned above. One of the main points
will be to elucidate the relation between those
probability distributions and 
the step fluctuations characterised by $G(t)$. Indeed, deviations
from the simple power law for $G(t)$ are found, which
may play a relevant role in interpreting experimental  
observations.\\

\noindent
The article is organized as follows. In the next section, the
model used to describe step fluctuations is introduced, and
the methods applied in the analysis, exact enumeration of random walks and 
Monte Carlo simulations, are introduced and outlined. Results
on the return probability distribution $p(l)$ are
given in section 3, followed
by a discussion of simulational findings on $G(t)$. A brief summary will
conclude the paper.\\
 
{\Large \bf 2. Model and methods}\\

Let us consider a (reference) surface step of monoatomic height with $2L+1$
sites, $i=-L,-L+1,...,0,...L-1,L$. The positions of the
step atoms, at time $t$,  are denoted by $h(i,t)$. Setting
the lattice spacing equal to one, $h$ is an integer. The step
is perfectly straight, if $h$ is constant. The neighbouring
terrace, consisting of sites $(i,j)$ on a square lattice, may be bordered
by a straight step opposite to the reference
step, separated by $d$ lattice spacings, see
Fig. 1. The opposite step, with $2L+1$ sites, is supposed to
either reflect (variant $s_r$) or absorb ($s_a$) an adatom diffusing
on the terrace, mimicing either a descending step with a
high Schwoebel--Ehrlich barrier or a rising step
with a large sticking coefficient.\\

\noindent 
Specific boundary conditions may be used at the ends of the reference
step, $i=-L$ and $i=L$. For instance, the ends may
be pinned or subject to periodic boundary conditions. Alternatively, the 
boundaries of the terrace perpendicular to the
reference and opposite steps may be, e.g., chosen
as straight reflecting (variant $t_r$) or absorbing ($t_a$)
straight steps with $d$ sites.\\

\noindent
Step fluctuations result from the detachment of an atom from the
reference step at site $i$, $h \longrightarrow h-1$, and attachment at
step site $i+l$, $h \longrightarrow h+1$. The rate of this process
depends on the energies involved and temperature. One may distinguish
three limiting cases, (a) $l=\pm 1$, corresponding to step diffusion;
(b) uncorrelated detachment and attachment sites, corresponding to
evaporation--condensation kinetics; and (c) initial and final sites
being connected by a random walk of the diffusing atom on the terrace
neighbouring the step, corresponding to terrace diffusion. Several
types of terrace diffusion are possible, reflecting various
constraints on the random walk, see below. The time dependence of the
step fluctuations may be quantified by the correlation function

\be
G(t)= \langle (h(i,t)- h(i,0))^2 \rangle
\ee

\noindent
The brackets denote a thermal average.\\

\noindent
We applied two methods in our study: (i) exact enumeration of random
walks [12], to investigate different kinds of terrace
diffusion, and (ii) Monte
Carlo simulations [13], to compute $G(t)$.\\

\noindent
(i) In enumerating random walks, we calculate the return
probability distribution $p(l)$ of an atom detaching
from site $i_0$ of the reference step onto
the terrace (the step positions being
then fixed once and for all, $h_0(i)= h(i,t)$) and attaching at step site
$i_0+l$. In other words, the atom starts its random walk
at terrace site $(i_0,h_0(i_0)+2)$, and one determines the
probability that it reaches the site
$(i_0+l, h_0(i_0+l)+1)$ first among all possible
absorption sites at the reference step. The terrace
is supposed to be bordered by either reflecting or
absorbing steps perpendicular and opposite to the
reference step. Obviously, one has  
 
\be
 \sum\limits_{l} p(l) \le 1 
\ee

\noindent
where the sum runs over all step sites. The identity holds when
the boundary steps are reflecting. Usually, the starting
point of the random walk is chosen at the center of the
step, $i_0= 0$.\\

\noindent
The exact enumeration of $p(l)$ is based on calculating the
probability $p_{\mathrm{rw}}(i,j,n)$ of the diffusing atom to be at
terrace site $(i,j)$ in the $n$-th move of the random walk [12,14].
The total probability, summing
$p_{\mathrm{rw}}(i,j,n)$ over all sites, is conserved
and equal
to one. $p_{\mathrm{rw}}$ is related to the probability to jump from the
neighbouring site $(i',j')$ to site $(i,j)$, $w(i',j',i,j)$, by
 
\be
 p_{\mathrm{rw}}(i,j,n)= \sum\limits_{i'j'} w(i',j',i,j) p_{\mathrm{rw}}(i',j',n-1) + w_s(i,j) p_{\mathrm{rw}}(i,j,n-1) 
\ee

\noindent
where $w_s(i,j)$ is the probability to stay at terrace site $(i,j)$. For
an absorption site, at the reference or a boundary step, one
has $w_s(ij)=1$ and $w(i,j,i',j')$=0, while for the
other terrace sites $w_s(i,j)=0$.\\

\noindent
On finite terraces, $p_{\mathrm{rw}}(i,j,n)$ can be calculated in a
straightforward fashion for a variety of jump probabilities and
boundary conditions, with the bookkeeping done by a computer programme
(exact analytic results are known only for rather few special cases
[15]). The return probability distribution $p(l)$ is given by
 
\be
p(l)= p_{\mathrm{rw}}(i_0+l,h_0(i_0+l)+1,n) 
\ee

\noindent
considering indefinitely long random walks, $n \to \infty$. In
practice, the length of the walk, $n$, depends on the convergence rate
of $p_{\mathrm{rw}}$. We studied steps of length $L \le 600$, with the
width of the terrace $d \le 600$.  Typically, the random walks ended
when the total probability of finding the atom on any non-absorption
terrace site was smaller than $10^{-8}$.\\

\noindent
(ii) In our Monte Carlo simulations an atom at step site $i$ is moved
to site $i+l$ with a (return) probability
$p_{\mathrm{da}}(l)$. The form of $p_{\mathrm{da}}(l)$  
is motivated by the findings on the random walks, $p(l)$.  The move is
then accepted, as usual, with a rate determined by the Boltzmann
factor of the associated energy change $\delta E$, $\exp(-{\delta
  E/k_BT})$, where $k_B$ is the Boltzmann constant and $T$ the
temperature [13]. The energies at the step are assumed to be given by
the number and depth of the kinks [16], as it is the case in the
standard SOS model [17] with the Hamiltonian
 
\be
{\cal H}=  \epsilon \sum_{[i,j]} \vert h(i,t) - h(j,t) \vert
\ee

\noindent
where the sum runs over neighbouring step
sites $[i,j]$, $j=i \pm 1$.-- To speed up the simulations, we
used an algorithm with a dynamic time assignment [13,18].\\ 
 
\noindent
The time $t$, elapsed during the simulation, is measured in terms of
Monte Carlo attempts per step site pair (MCA), i.e.  in one time unit
one has tried to interchange, on average, one atom between each two
step sites (the Monte Carlo time scale is linearly related to the real
time).  This interpretation corresponds to the situation where the
time spent by the diffusing atom on the terrace is negligibly small
compared to the mean time spent at the step. More realistic approaches
using, e.g., kinetic Monte Carlo simulations for the surface dynamics,
may suffer from other shortcomings, such as an ambiguity in the step
position. Note that the average position of the entire step, $\sum
\limits_i h(i,t) / (2L+1)$, does obviously not depend
on time.\\

\noindent
To compute the step fluctuations $G(t)$, see eq. (1), one
has to define a starting time, $t=0$. Various choices
are possible, corresponding to various initial step
configurations, including straight and equilibrated steps.
Perhaps closest to experiments, one may average over
an ensemble of thermalized
step configurations, possibly during
a single Monte Carlo run (which will be called, in the
following, dynamic averaging). In addition, one may
impose different boundary conditions, for instance,
by pinning the end positions of the step, at $i=1$ and $2L+1$,
or applying periodic boundary conditions. In the
later case, boundary effects are less severe.\\

\noindent
Typically, we studied steps of length $2L+1 \le 256$, with
periodic boundary conditions.\\

{\Large \bf 3. Random walks}\\

Using the exact enumeration approach, the return probability 
distribution $p(l)$ has
been calculated for four distinct cases, mimicing possible
constraints on the terrace diffusion: (a) non--perturbed random
walk, $w(i',j',i,j)= 1/4$, with a straight reference
step, see Fig. 2; (b) diffusing
atom in an external potential, $V= A/y^2$, $y$ being the
distance between the (straight) steps and the atom, due to, e.g., elastic
interactions [19], see Fig.3. Note that
due to the potential $V$ the jump
probability $w \propto \exp (-\delta V/k_BT)$,
 with $\delta V= V(i',j')- V(i,j)$, becomes anisotropic, favouring
hops away from the step; (c) unbiased
diffusion, $w= 1/4$, on a
terrace with quenched absorbing single--site
defects of concentration $c$, with
a straight reference step, see Fig.4; and (d) non--perturbed random
walk with a rough reference step (the roughness mimics
a thermalized reference step).\\

\noindent
In each case, the wandering atom started at the center of the
reference step, at terrace site $(0,h_0(0)+2)$. The three boundary
steps were assumed to be straight and either reflecting or
absorbing. For instance, '$s_r:t_a$' refers to a situation with
a reflecting opposite step and absorbing terrace boundary steps
perpendicular to it.\\

\noindent
In general, one may distinguish four regimes, in which $p(l)$
exhibits different characteristic properties, depending on the
return distance $l$. At very short distances of a few lattice
spacings, $l < l_0$,  $p(l)$ falls off rapidly, the concrete form being
determined by
details of the perturbations (step roughness, concentration
of defects and strength $A$ of the elastic interaction).\\

\noindent
At $l_0 \ll l \ll d$, $p(l)$ acquires typically a power--law
behaviour, $p(l) \propto l^{ \alpha}$. To analyse that regime, one
may calculate the effective exponent 

\be
\alpha_{\mathrm{eff}}(l)= d \ln p(l)/ d \ln l 
\ee

\noindent
being constant, $\alpha_{\mathrm{eff}}= \alpha$, if, indeed, the deacy
of $p(l)$ follows a simple power--law. Increasing the return distance
$l$ furthermore, the effect of the boundaries on the random walk shows
up. Assuming $L \gg d$, first the presence of the opposite (vicinal)
step affects the return probability, leading to a characteristic
exponential decay of $p(l)$, $p \propto \exp (-al)$, both for
reflecting, $s_r$, and absorbing, $s_a$, steps.  Finally, the
distribution $p(l)$ will be modified due
to the perpendicular terrace boundaries, $t_r$ or $t_a$.\\

\noindent
As illustrated in Fig. 2 for the non--perturbed case,
$\alpha_{\mathrm{eff}}$ tends to approach $\alpha= -2$, in agreement
with the solution of the corresponding continuum diffusion equation
[9]. Consequently, one expects $p(l)=p_0 l^{-2}$ at sufficiently large
return distances $l$ for indefinitely long isolated straight steps
(note that $\alpha$ is expected to determine the
value of the dynamic exponent
$\gamma$ describing the time dependence of the step fluctuations
$G(t)$ [9], see below).  Actually, neither the interaction of the
diffusing atom with the steps nor the step roughness seem to affect
that value of $\alpha$. In fact, for a rough reference step,
$\alpha_{\mathrm{eff}}(l)$ follows closely the form for a straight
step, $l > l_0$ [14]. Applying the potential $V= A/y^2$, we observe a
systematic dependence of the proportionality factor $p_0$ on A, as
shown in Fig.  3. $p_0$ is found to increase exponentially with the
interaction strength $A$, $p_0 \propto \exp (\eta A)$, with $\eta
\approx 1.23$, at least for the strengths we considered, $0 \le A \le
1$. As depicted in Fig. 4, our findings do not rule out that the value
of $\alpha$ may, however, depend on the concentration $c$ of defects
on the terrace. Over an appreciable range of return distances $l$ one
may notice a plateau--like behaviour in $\alpha_{\mathrm{eff}}$ at
a value slightly,
but definitely smaller than --2. E.g., at $c=0.2$, the
plateau--like behaviour occurs
at about $-2.2$; at $c= 0.1$, the effective
exponent $\alpha_{\mathrm{eff}}$ seems to settle for some distances
$l$ at a value below, but closer to --2. However, the
data do not allow to rule out the possibility that $\alpha= -2$ may
be approached, now from below, at larger return distances $l$. Of 
course, it would be desirable to consider longer
steps and larger terraces to study further this aspect. Such
computations would be extremely computer time consuming, because one
has to average over a large number of defect concentrations to get
meaningful data (typically, we
averaged over about 10 000 realizations).\\

\noindent
With increasing $l$, the return probability distribution $p(l)$ of the 
diffusing atom approaches an exponential decay, $p(l) \propto \exp(-gl)$ in
all four cases, $(a)$--$(d)$. Accordingly, the exponential form describes 
$p(l)$ for large return distances $l$ in the presence of
a neighbouring, descending or rising, step, separated by
$d$ lattice spacings. For the unbiased random walk with
straight steps the coefficient $g$, as
follows from exact enumeration, agrees, to a high degree
of accuracy, with the one obtained from the continuum
diffusion equation [9], namely $g= \pi /d$ for an absorbing
opposite step, and $g= \pi /2d$ for a reflecting, vicinal step. The
various perturbations seem to have little effect: The
decay form of $p(l)$ remains exponential, and even
the coefficient $g$ seems to be rather robust.-- In the unbiased
case, $(a)$, we found that $p(l,d)$ seems to scale, at $d \gg 1$,
as $p(l,d)= p_s(l/d)/d^2$, both in the power--law and
the exponential regime [14].\\

\noindent
Finally, with $l$ getting closer to $L$, the perpendicular terrace
boundaries become relevant. As expected, $p(l)$ is lower when the
boundaries are absorbing than in the reflecting situation. However, we
did not explore this region in much detail, because those 
boundary conditions are somewhat artificial and extremely
long random walks, see eq. (4), would be needed for quantitatively
correct results.\\

{\Large \bf 4. Monte Carlo simulations}\\

We simulated the fluctuations $G(t)$ of steps of length $M=2L+1$,
with their ends being connected by periodic boundary conditions. An
atom is detached at the randomly chosen site $i$ and attached
at site $i+l$, see Fig. 1. The (return) probability
of selecting site $i+l$ is denoted by $p_{\mathrm{da}}(l)$. The 
probability of accepting such an elementary move is given
by the Boltzmann factor of the energy change, see eq. (5), associated
with this process.\\ 

\noindent
We studied mainly five distinct cases, with the choice of
$p_{\mathrm{da}}(l)$ being motivated by results on the return
probability distribution $p(l)$ of a random walk and by general physical
considerations.  In particular, we considered $p_{\mathrm{da}}(l)$
being (i) constant, corresponding to evaporation--condensation
kinetics.  Obviously, the mean step position is conserved in each
move.  We also simulated evaporation--condensation dynamics by
choosing between detachment and attachment completely randomly, thus
conserving the mean step position only in the time average; (ii) 1/2
for $l= \pm 1$ and 0 elsewhere, describing step diffusion; (iii)
proportional to $1/l^2$ (the prefactor is determined by requiring
that $p_{\mathrm{da}}(l)$ is normalized: $\sum p_{\mathrm{da}}(l) =
1$), as one may expect for fluctuations of an isolated step due to
terrace diffusion, neglecting deviations from that form at short
distances $l$ between attachment and detachment sites; (iv)
proportional to $\exp (-l)$, monitoring the possible effect of a
neighbouring step on $G(t)$; and (v) proportional to $1/l^3$, to study
the sensitivity of $G(t)$ on the value of $\alpha$ in
$p_{\mathrm{da}}(l) \propto l^ {\alpha}$.\\

\noindent
For indefinitely long steps, one may argue that $G(t)$ increases
with time in the form of
a power--law $G(t) \propto t^{ \gamma}$ [7-10], with the dynamic exponent
$\gamma$ depending on the atomistic mechanism governing
the fluctuations. To determine $\gamma$ in the various cases we
simulated, one may calculate the effective dynamic exponent

\be
\gamma_{\mathrm{eff}}(t)= d \ln G(t)/ d \ln t 
\ee

\noindent
with $\gamma_{\mathrm{eff}}(t)$ approaching $\gamma$ at large times and long
steps.\\

\noindent
Results on $\gamma_{\mathrm{eff}}(t)$ for each of the five situations
are depicted in Fig. 5, at fixed step length, $M= 256$, and fixed
temperature $k_BT/ \epsilon =1$. We used dynamic averaging over
an ensemble of successive, equilibrated initial
configurations. Evaporation--condensation kinetics was simulated
by random attachment and detachment processes, i.e. without
conserving the average step position in each Monte Carlo move.\\

\noindent
At very early times, the step fluctuations are always diffusive, with
$G(t) =c_d t$, corresponding to $\gamma_{\mathrm{eff}}= 1$. The diffusion
coefficient $c_d$ depends rather weakly on the transport mechanism, with
the exception of the evaporation--condensation dynamics having
a significantly larger diffusion coefficient.\\

\noindent
Due to the rigidity of the step, the step meandering then slows down,
with $\gamma_{\mathrm{eff}}$ approaching a plateau located at about 1/2, 1/3,
or 1/4, depending on the form of the return probability,
$p_{\mathrm{da}}$, see Fig. 5. Actually, the plateau at $\gamma_{\mathrm{eff}}
\approx 1/2$, characterising evaporation--condensation, is reached
most quickly. The plateau at 1/3 is realised for terrace diffusion
with an isolated step, $p_{\mathrm{da}}(l) \propto 1/l^2$. The largest
time is needed to approach the plateau at 1/4, characterising step
diffusion as well as terrace diffusion with rapidly decaying return
probabilities, $p_{\mathrm{da}}(l) \propto 1/l^3$ and $\propto \exp
(-l)$. The last situation corresponds to terrace diffusion with pairs
of steps. One may argue that the plateau signals the asymptotic power
law increase of $G(t) \propto t^{\gamma}$ for large times and
indefinitely long steps. Indeed, the plateau values have been obtained
before in Langevin descriptions [7-9] for infinitely extended steps
(for $ -3 < \alpha < -2$, one may expect a continuously
varying value,  $\gamma= 1/ (-\alpha +1)$ [9]).\\

\noindent
Combining our results on the random walks and simulations, we conclude
that the exponent $\gamma= 1/3$ for an isolated
step is robust against
various perturbations of terrace diffusion, with the possible exception of
defects on the terrace.\\

\noindent
The estimates of the plateau values of $\gamma_{\mathrm{eff}}$ are
confirmed by a Fourier analysis of the step fluctuations $h(i,t)$.
They can be written in the form

\be
h(i,t)=  \sum\limits_{k=1}^{M/2} (a_k(t) \sin (2\pi k i/M) + b_k \cos (2\pi k i/M)) 
\ee

\noindent
The fluctuation modes may be described by

\be
G_k(t)= \langle (a_k(t)- a_k(0))^2 + (b_k(t) -b_k(0))^2 \rangle
\ee

\noindent
taking into account phase shifts. Based on
Langevin descriptions, $G_k(t)$ is expected to converge
rapidly towards equilibrium $G_k (\infty)$ [7-9],

\be
G_k(t)= G_k(\infty) (1- \exp(-I_k t)) 
\ee

\noindent
in agreement with the simulational data. More specificly, for 
small wavenumbers, $I_k$ 
follows closely the form $I_k \propto k^x$, with $x= 1/\gamma$, in
accordance with the Langevin theory. However, $I_k$ shows
pronounced deviations from the power--law behaviour at larger
values of $k$, as shown in Fig. 6. This feature is not described
by the Langevin theory [7].\\  

\noindent
At later times, after having passed the plateau, the effective
dynamic exponent, $\gamma_{\mathrm{eff}}$, will eventually go to
zero (as may be easily seen for shorter steps [11]) or, in the
case of evaporation--condensation, to
one, due to the constraint of conserving, or not, the average
step position, see Fig. 5. For evaporation--condensation, the
related diffusion
coefficient is rather small and depends on the step length. In 
any event, the late--time behaviour
reflects the finite length of the step.\\

\noindent
The time dependence of $\gamma_{\mathrm{eff}}$, as depicted in Fig. 5,
implies that an average effective exponent, $\gamma _a$, as obtained
from a log--log plot of $G(t)$ in a given time range, may vary with
experimental parameters when one is not in the truely asymptotic
regime. In particular, the plateau in $\gamma_{\mathrm{eff}}$ may not
yet have been reached or finite size effects may already matter. Note
that, for instance, the crossover from the diffusive short--time
behaviour to the subdiffusive step motion at later times is
temperature dependent, thereby causing possibly a temperature
dependence in $\gamma_a$. Similarly, the presence of a neighbouring
step is expected to affect $\gamma_a$. In general, much care is needed in
identifying and disentangling the numerous possible crossover effects.\\

{\Large \bf 5. Summary}\\

We studied equilibrium step fluctuations
in a somewhat idealized way by enumerating
random walks on terraces with various constraints and using
Monte Carlo techniques.\\

\noindent
In the simulations, different atomistic mechanisms
determining the equilibrium step fluctuations may be mimiced by special
choices of the return probability for atoms attaching
at distance $l$ from the detachment site of the step. 
The form of the return probability at large distances $l$
determines the time dependence of the step fluctuations $G(t)$
at late times $t$ for long steps. In particular, we simulated
evaporation--condensation kinetics, step diffusion and (perfect) terrace
diffusion. For ${\it isolated}$ steps, $G(t)$
is confirmed to increase
with a power law $G(t) \propto t^{\gamma}$, with $\gamma$ =1/2, 1/3, and 
1/4, respectively. In the case of terrace diffusion 
for ${\it pairs}$ of steps, the dynamic exponent $\gamma$
approaches the value of step
diffusion, $\gamma =1/4$. The simulational observations 
agree with and refine predictions of Langevin theory.-- Similarly, a
Fourier analysis of the step
fluctuations, driven by the different atomic mechanisms, confirms and
refines previous Langevin descriptions.\\

\noindent
In general, crossover phenomena may mask the
asymptotic behaviour of $G(t)$. Such phenomena can be caused by
several reasons, including change
from diffusive to
subdiffusive step motions at early times, effect
of the finite step length, defects on the terrace, and influence
of neighbouring steps. As a result, the
average effective dynamic exponent $\gamma_a$, as usually
obtained from measurements, may vary with experimental
parameters like temperature.\\ 

\noindent
From the exact enumeration of random walks, the 
decay of the return probability distribution $p(l) \propto l^{-2}$
for unbiased terrace diffusion with straight steps is found
to be robust against step roughness and (elastic)
interactions between the diffusing atom and the neighbouring
step. This fact implies the robustness of the value $\gamma= 1/3$
for isolated steps. Moreover, for pairs of steps, the
exponential decay of $p(l)$ persists in the presence
of those perturbations, implying the robustness of $\gamma= 1/4$
in that case.\\
 
{\bf  Acknowledgement}

\noindent
We should like to thank H. P. Bonzel, T. L. Einstein, H. Ibach, and
M. Giesen for useful discussions and remarks.\\

\vskip 1.3cm
\newpage
\bc
{\Large \bf References}\\[2ex]
\ec
\begin{enumerate}
\item L. Kuipers, M. S. Hoogeman, and J. W. M. Frenken,  Phys.
Rev. Lett. 71 (1993) 3517.
\item M. Giesen-Seibert, R. Jentjes, M. Poensgen, and H. Ibach,  Phys.
Rev. Lett. 71 (1993) 3521.
\item M. Giesen, G. S. Schulze Icking--Konert, D. Stapel, and H. Ibach, Surf.
Sci. 366 (1996) 229.
\item W. W. Pai, N. C. Bartelt, and J. E. Reutt--Robey, Phys. Rev. B 53 
(1996) 15991.
\item P. Wang, H. Pfn\"ur, S. V. Khare, T. L. Einstein, E. D. 
 Williams, W. W. Pai, and J. E. Reutt--Robey, Phys. Rev. B 53 
(1996) 15991.
\item S. Tanaka, N. C. Bartelt, C. C. Umbach, R. M. Tromp, 
 and J. M. Blakely, Phys. Rev. Lett. 78 (1997) 3342. 
\item N. C. Bartelt, T. L. Einstein, and E. D. 
 Williams, Surf. Sci. 312 (1994) 411. 
\item A. Pimpinelli, J. Villain, D. E. Wolf, J. J. Metois, J. C.
 Heyraud, I. Elkinani, and G. Uimin, Surf. Sci. 295 (1993) 143. 
\item B. Blagojevic and P. M. Duxbury, in : Dynamics of Crystal
Surfaces and Interfaces, Eds. P. M. Duxbury and T. Pence (Plenum, New York,
 1997) p.1. 
\item S. V. Khare and T. L. Einstein, Phys Rev. B57 (1998) 4782. 
\item W. Selke and M. Bisani, Lecture Notes in Physics 519 (1999) 298. 
\item I. Majid, D. Ben-Avraham, S. Havlin, and H. E. Stanley, Phys
Rev. B30 (1984) 1626. 
\item K. Binder and D. W. Heermann, Monte Carlo Simulations in Statistical
Physics (Springer, Heidelberg, 1992). 
\item M. Bisani, Diploma thesis, RWTH Aachen (1998). 
\item W. Feller, An Introduction to Probability Theory and its 
Applications (John Wiley, New York, 1968). 
\item B. S. Swartzentruber, Y.-W. Mo, R. Kariotis, M. G. Lagally,
 and M. B. Webb, Phys. Rev. Lett. 65 (1990) 1913. 
\item J. D. Weeks, in : Ordering in Strongly Fluctuating Condensed
Matter Systems, Ed. T. Riste (Plenum, New York, 1980) p.293. 
\item A. B. Bortz, M. H. Kalos, and J. L. Lebowitz, J. Comp. Phys. 17 (1975) 10. 
\item V. I. Marchenko and A. Y. Parshin, Sov. Phys. JETP 52 (1980) 129. 
\item W. H. Press, S. A. Teukolsky, W. T. Vetterling, and B. P. Flannery, Numerical Recipes in C (Cambridge University Press, Cambridge, 1992). 
\end{enumerate}

\newpage
\bc
{\Large \bf Figure Captions}\\[2ex]
\ec
\begin{itemize}
\item[Fig. 1:] Geometry of the surface, showing the reference step
and a descending ($s_r$) or rising ($s_a$) opposite step.
 
\item[Fig. 2:] Effective exponent $\alpha_{\mathrm{eff}}$ of the
  return probability distribution $p(l)$ for the
  unbiased random walk with a straight reference step
  ($L=600$, $d=600$). Absorbing terrace boundaries, $t_a$, 
  were used.
 
\item[Fig. 3:] Return probability distribution $p(l)$ for random walks
  in a repulsive external potential $V = A y^{-2}$ at $k_BT=1$, for $L=600$,
  $d=600$, and with $s_a:t_a$ boundary conditions.
 
\item[Fig. 4:] Effective exponent $\alpha_{\mathrm{eff}}$ of
  $p(l)$ for
  random walks with defects on the terrace ($d = 50$, $L=100$,
  concentration of defects $c=0.2$, averaged over 8778 realisations,
  with $s_a:t_a$ boundary conditions). The dashed line is based on smoothed
  data obtained by applying a Savitzky-Golay-filter [20].
  
\item[Fig. 5:] Effective dynamic exponent $\gamma_{\mathrm{eff}}$ as
  a function of time (measured in units of MCA) for various atomistic
  mechanisms and return probabilities.  From top to bottom:
  evaporation--condensation
  (solid), $p_{\mathrm{da}}(l) \propto l^{-2}$ (solid), 
  $\propto l^{-3}$ (solid), $\propto e^{-l}$ (dotted),
  $=1/2$ for $l=\pm 1$ (dashed).  Data was smoothed using a
  Savitzky-Golay-filter. Steps with 256 sites, at
  $k_BT/\epsilon= 1$ were simulated.
  
\item[Fig. 6:] $I_k$ vs. $k/M$ (in units of $2 \pi$) 
  for different transport
  mechanisms: evaporation--condensation (plus), terrace diffusion with
  $p_{\mathrm{da}}(l) \propto l^{-2}$
  (cross) and step diffusion with $p_{\mathrm{da}}(l)=1/2$ for
  $l=\pm 1$ (asterik).  The
  lines correspond to the power--law behaviour expected from Langevin
  theory: $I_k \propto k^2$ (solid), $I_k \propto k^3$ (dashed) and
  $I_k \propto k^4$ (dotted), respectively. Steps with 64 sites, at
  $k_BT/\epsilon= 0.8$, were simulated.

\end{itemize}
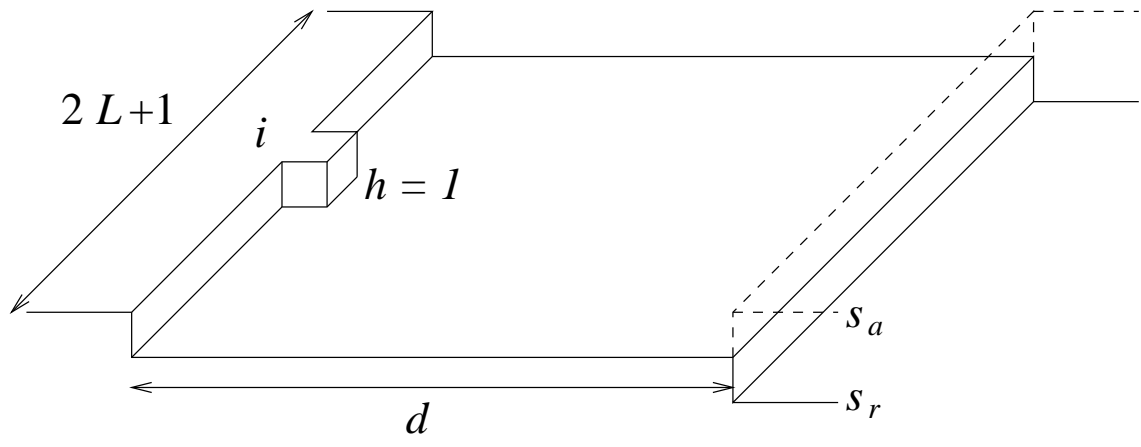
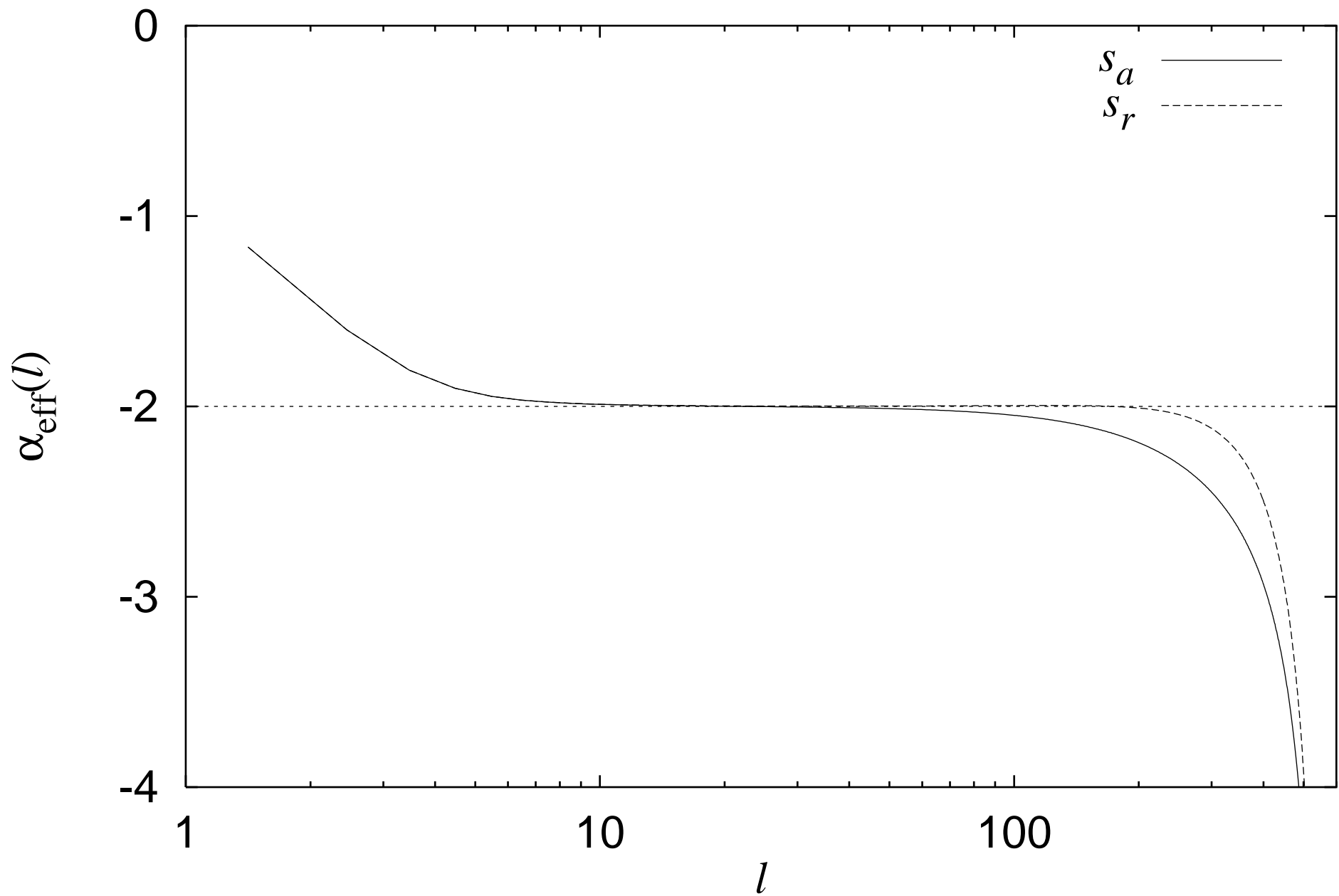
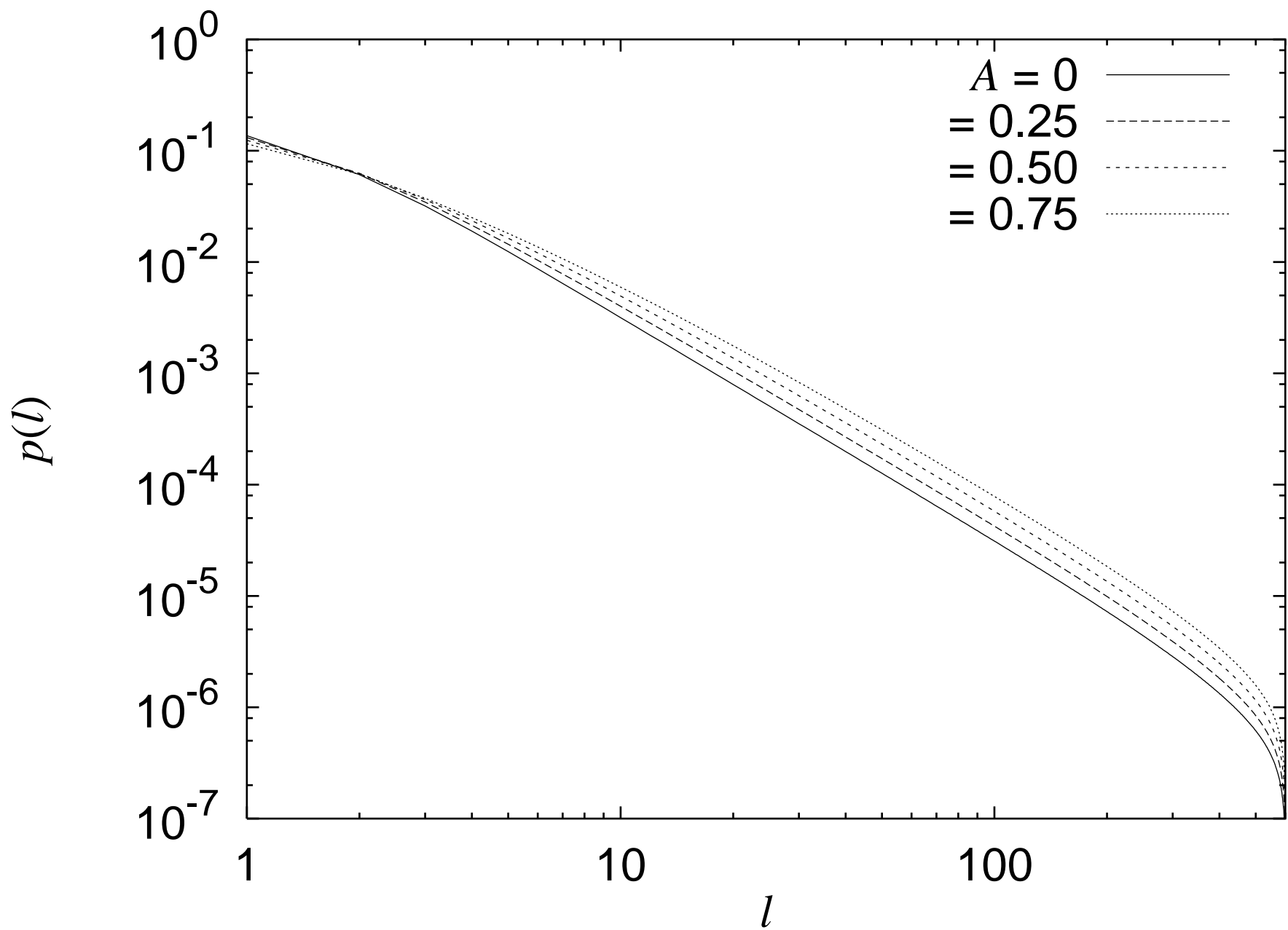
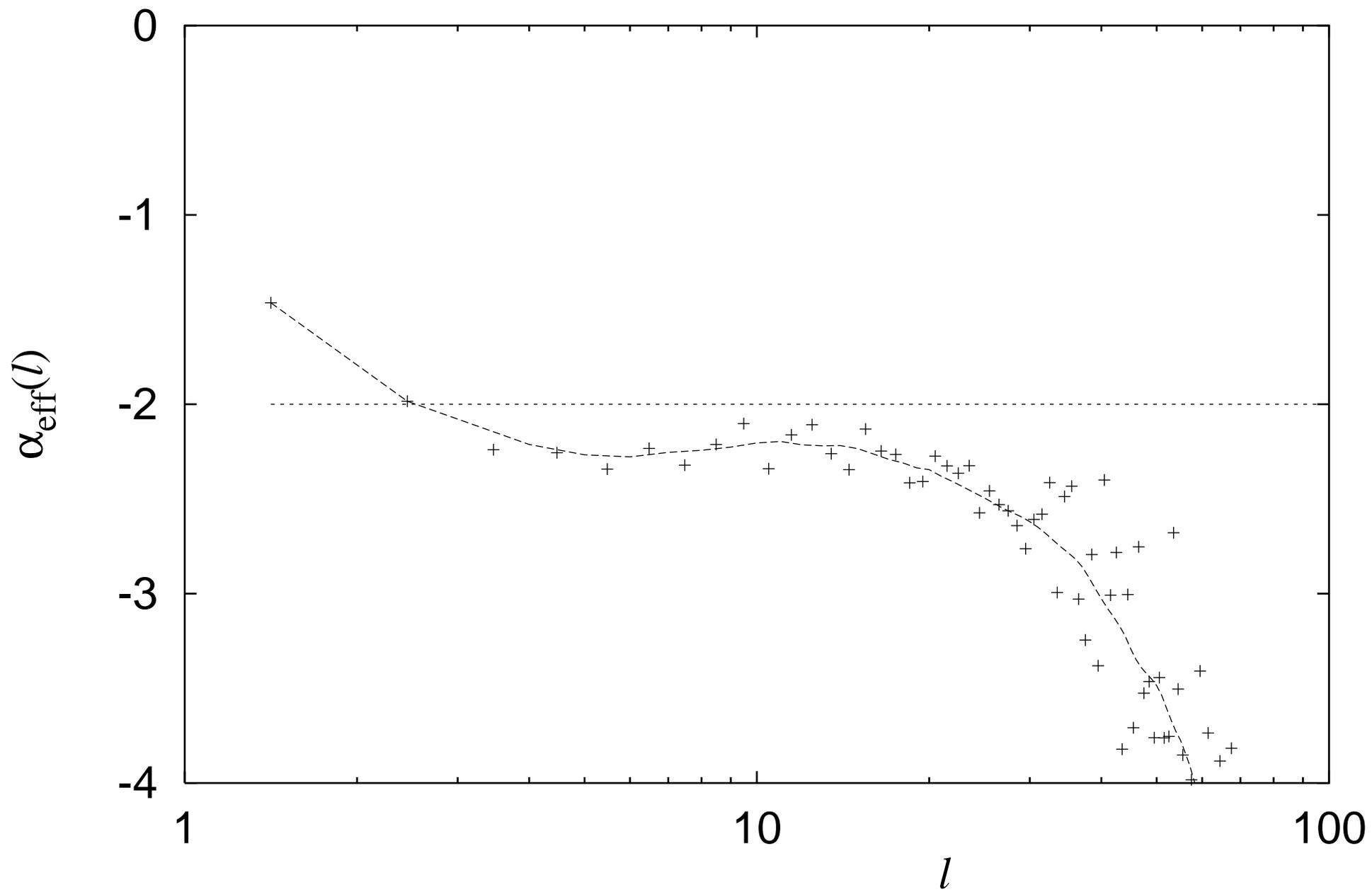
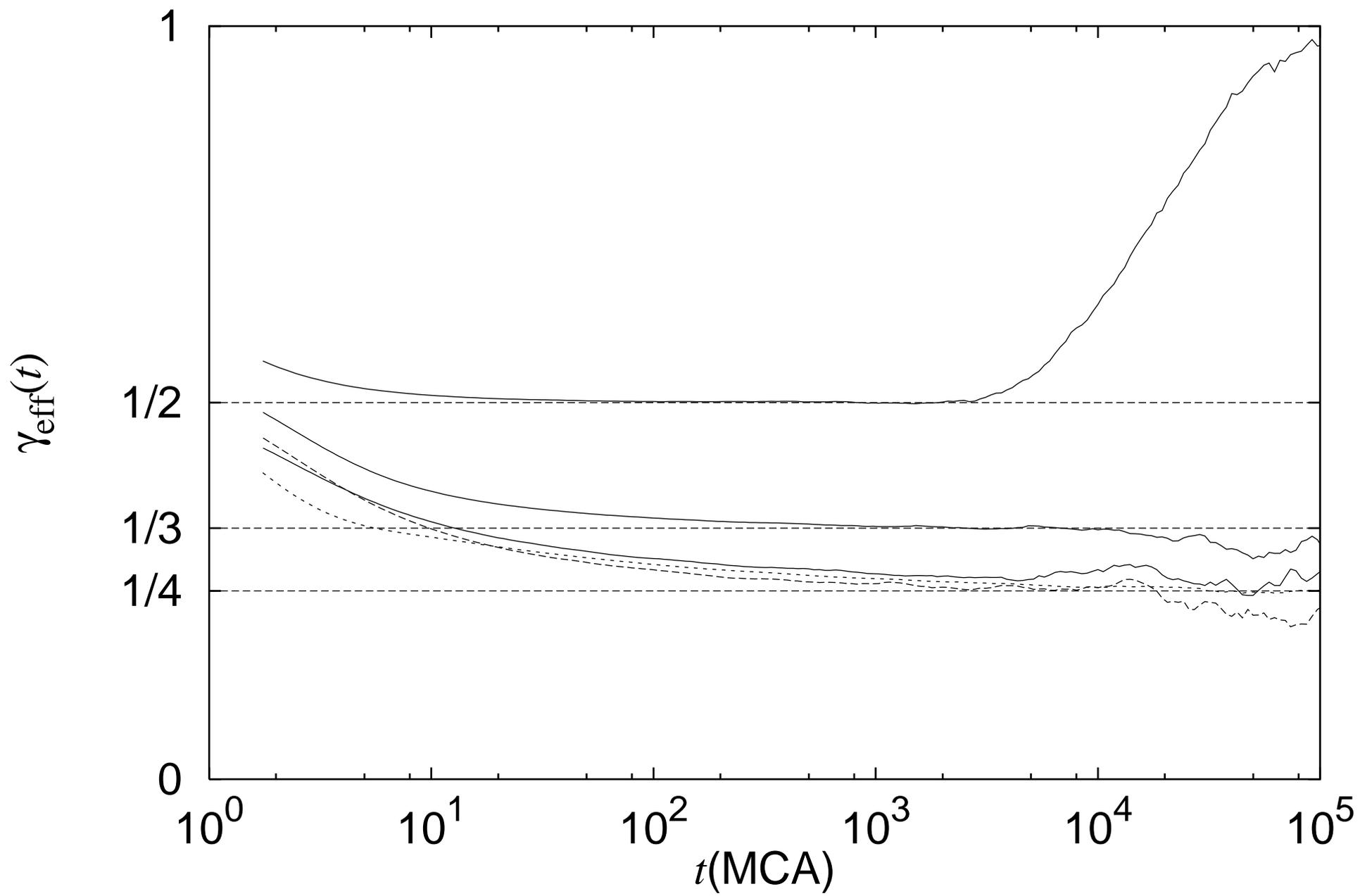
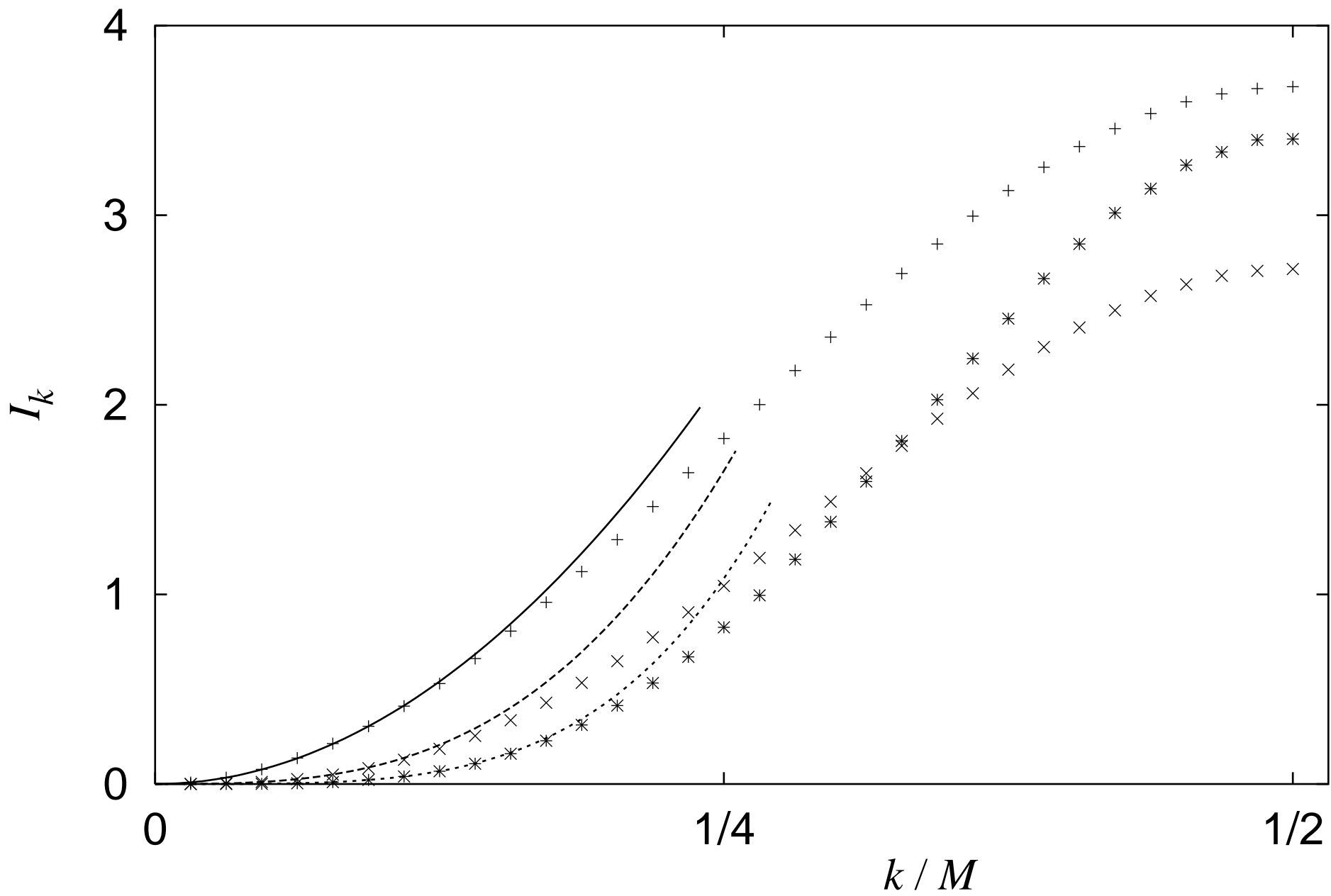
\end{document}